# INFOSEC COMPETITION 2014

# ACADEMIC RESEARCH PAPER

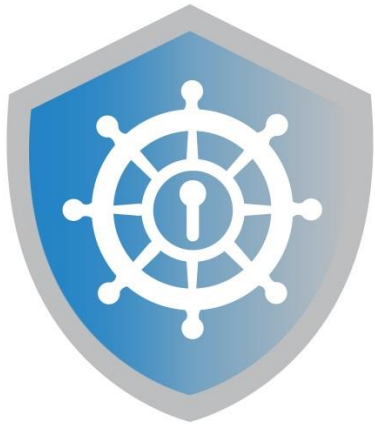

CAS-SUR IT Security Group

| Project Name: | **INFORMATION SECURITY AWARENESS AT OMAN EDUCATIONAL INSTITUTIONS: AN ACADEMIC PERSPECTIVE** |
|---|---|
| Team Name: | CAS-SUR-ITSecurity |

**Project Members**

| Designation | Name | Email | Contact Number |
|---|---|---|---|
| Coordinator | Mr. Rajasekar Ramalingam | rajasekar.sur@cas.edu.om | 92229686 |
| Member | Dr. Ramkumar Lakshminarayanan | ramkumar.sur@cas.edu.om | 92912903 |
| Member | Mr. Shimaz Khan | shimaz.sur@cas.edu.om | 92331022 |

Ministry of Higher Education
Sur College of Applied Sciences
Department of Information Technology
Post Box: 484 Post Code: 411
Sultanate of Oman

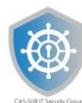

# TABLE OF CONTENT

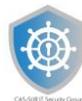

# CHAPTER 1

## PROJECT DESCRIPTION

### 1.1 Introduction

In the late 1990s, after the significant disruption caused by the computer viruses such as Melissa and Code Red the need for information security received attention globally.  Since then there has been many new threats such as spam emails [1], identity theft [3], data leakage [7], phishing [8], adware [9], intrusion [10] and many more are evolving continuously and have considerable impact on the information assets of organization and individual.

### 1.2 Internet usage in Oman

On the other hand, according to the World Internet usage statistics news [2], Internet users in the Sultanate of Oman have been continuously increasing in recent days. Oman is noteworthy for marketers due to their excellent consumer protective laws making online purchases a safer one and boost up the online buying of the individuals [4].  Steady economic growth has increased the volume of the net banking, mobile banking and improvements in payment infrastructure have driven the growth of Oman's card payments channel.  The card payments channel is forecasted the growth of card usage will be around 4.4 million by 2017 [5]. The high number of internet usage, increased card usage, and the revolution of internet technology has led a significant increase in the number of online transactions and electronic data transfer.  At the same time the number of cybercrime incidents in Oman is also increased. According to the annual report of Information Technology Authority (ITA) (2012 and 2013) [6], there is a significant increase in the number of cybercrime incidents in Oman.

### 1.3 IT Security incidents in Oman

As per the ITA annual report (2012 and 2013), the number of malicious attempts against secure government portals is increased.  Compared to 2011, there was an increase of 13.5% reported incidents and 200% increase of malware incidents.  1,084,369 malicious attempts were prevented and analyzed against government portals. 19,171 malicious attempts against government networks were identified and prevented. 25,827 vulnerabilities were discovered by scanning 9,890 IPs. 941,079 malicious-ware were analyzed to determine the main cause of infection. 659,090 web violations were analyzed and prevented. 15,855 security attacks were discovered and handled by OCERT.

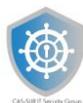

**1.4 Problem definition**

Attackers have adopted stealthier techniques to exploit user trust, this strategy leads to the most variable, unpredictable and critical vulnerability to the information assets. The gap between the knowledge and the practices regarding the computer and Internet use are the major factors which influence security of Information.

In this academic research paper a survey was performed on the education institutions in Oman to investigate the level of information security awareness among various entities which includes students, technical staff and academic staff. The survey focused on the following factors of Information Security: 1) Demographics, 2) Internet usage, 3) Awareness about organization's network 4) Security threats experience, 5) Awareness on Password management, 6) Email security awareness and 7) Knowledge on security practices. The survey attracted 173 respondents: the level of information security awareness and knowledge on security practices were correlated and analyzed. The areas of weakness related to the security implementation, awareness, and practice were identified.

**1.5 Methodology**

A proposed model, Information Security Awareness Identification Model (ISAIM) is adopted in this project. As shown in Figure 1.1, ISAIM has 6 key elements which include 1) Effective usage, 2) Organization awareness, 3) Threats awareness, 4) Protection awareness, 5) Content awareness and 6) Security practice.

- **Effective usage** element identifies how frequent the user use IT enabled devices, the purpose of usage and access location.

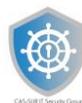

- **Organization awareness** element identifies the user's knowledge about their environment like availability of IT infrastructure, security policy, and security standards.

- **Threats awareness** element refers to finding the threat experience of an individual, the frequency, loss due to attacks, knowledge of security policy and reporting mechanism.

- **Protection awareness** element identifies how efficiently identities of the individuals choose and manage.

- **Content awareness** element identifies how the user evaluates the validity of a email content.

- **Security practice** element identifies the necessary components of security habits includes skills needed, training attended and training required.

- A survey of Oman educational institutions responds will assist in determining the awareness importance measure.

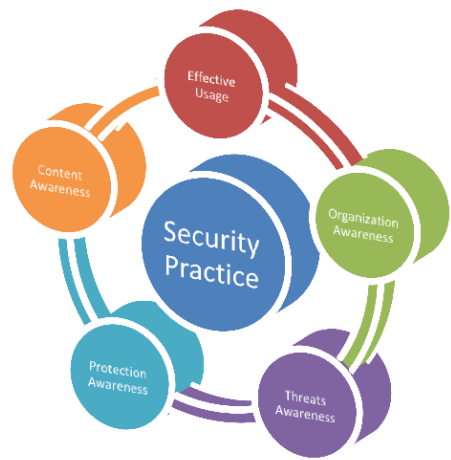

**Figure 1.1: ISAIM Model**

**1.6 Target group and reasons for choosing**

The target group is the entities of the educational institutions in Oman which includes the academic staff, technical staff and students.  The reasons for choosing the group are listed below:

Academic staffs are the main pillars of educational institutions; they are playing a vital role in articulating the technical knowledge of the students and major user of Information

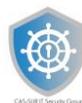

Technology.  It is essential to find the level of security awareness that exist among the academic staff community.  The purpose of finding the level of security awareness among the technical staff is to find how they are using technology to protect the information assets of the institution.

| S# | Name of educational institution |
|----|--------------------------------|
| 1 | Al Buraimi University College |
| 2 | Higher College of Technology |
| 3 | Ibra College of Technology |
| 4 | Salalah College of Technology |
| 5 | Sur College of Applied Sciences |
| 6 | Waljat College of Applied Sciences |
| 7 | Sohar College of Applied Sciences |
| 8 | Nizwa College of Technology |
| 9 | Oman College of Management Technology |
| 10 | Al Sharqiyah University |
| 11 | German University of Technology in Oman |
| 12 | Ibri College of Applied Sciences |
| 13 | Majan University College |
| 14 | College of Applied Sciences, Rustaq |
| 15 | Sultan Qaboos University |
| 16 | Caledonian College of Engineering |
| 17 | Sohar University |

**Table 1.1 List of Oman educational institutions participated in survey**

Students are our ambassadors and they are the one, who is going to implement all the skills and knowledge learned in the various organizations of Oman in the near future.  They must know how to secure their work environment and the information they are dealing with. They do require a detailed understanding of the security threats, damages by various threats and solutions to mitigate the damage.  Table 1.1 shows the list of educational institutions that participated in the survey.

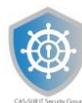

# CHAPTER 2

## PROJECT IMPLEMENTATION

This section describes the implementation of ISAIM methodology to find the level of Information security awareness. A survey was conducted across the educational institutions in Oman to identify the security practice elements of the model and leverage the opinions of entities of the educational institutions. 173 respondents' data were collected and analyzed: the key findings are as follows.

**Survey URL: https://www.surveymonkey.com/s/P69T3SV**

### 2.1 Effective usage

| Item | Description |
|------|-------------|
| Target group | Students, Technical staff and Academic Staff |
| Topics covered | Gender, age group, educational qualification, current position, device to access internet, way of accessing Internet, purpose and duration. |
| **Key findings** ||

- Among 173 respondents, 76.9% were male and 21.3% were female.
- The age group of the respondents were scattered and a considerable percentage of respondents (34%) falls in the category of 18 to 29.
- 35% of graduates, 38% of masters, and 23% of PhD holders were participated in the survey.
- 54% of respondents were academic staff.
- Majority of the users use the smart phone device (70%) to access the internet, a least number uses tablets (17%).
- Majority of the responded accessing the internet from home. The second major internet access location is the University / College / Institute.
- Only 7% of the responded are using internet cafes for accessing the internet.
- Emailing and educational references are the major purpose for using internet, while the

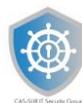

social networking and net banking hold the next highest need for using the internet.

- 27% of the responded use internet on an average of 2 to 3 hours and day and 14% of responded use more than 7 hours a day.

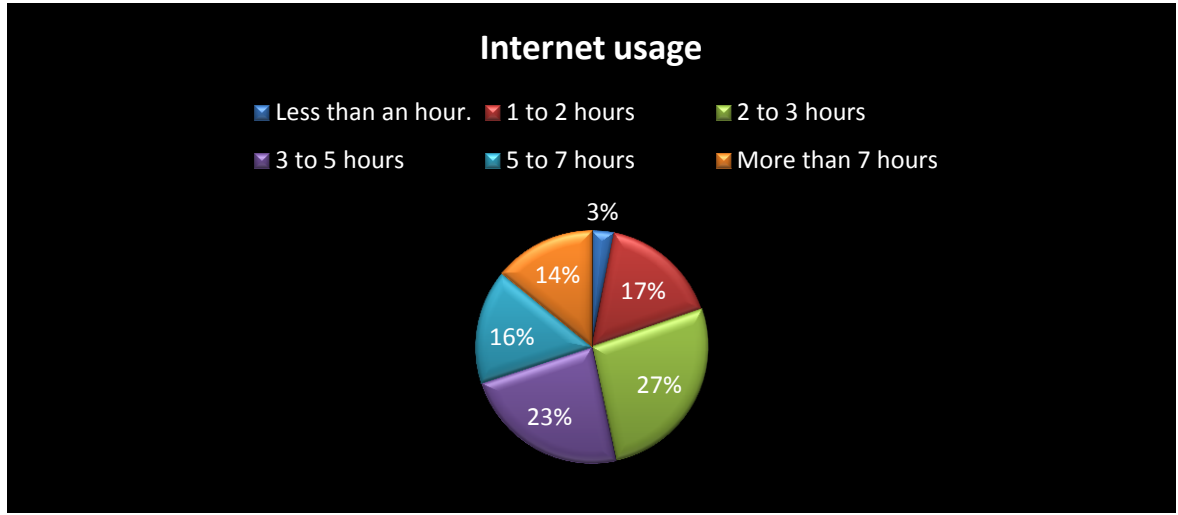

## 2.2 Organization awareness

| Item | Description |
|------|-------------|
| Target group | Student, Technical Staff and Academic Staff |
| Topics covered | Awareness on Information security management system standard, Local Firewall, Intrusion Detection Systems, Demilitarized Zone, Antivirus software and Security software used to protect system. |
| **Key findings** | |

- It is very surprising to know from the survey that 22% of the respondents reveled that there is no information security standard practices in their institutions and 39% responded that, they were not aware about such practices.
- The top most security technology used in the educational institutions in Oman is Antivirus software and Firewall.
- The majority of the educational institutions use Firewall as their security tool to protect their network.

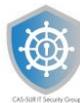

- The survey reveals that the Intrusion Detection System and DMZ are still a mysterious term to the respondents, 49% were not aware whether IDS is deployed or not and 61% were not aware about the DMZ.
- 2.6% of respondents gave the response "I don't use any security software".

### 2.3 Threats awareness

| Item | Description |
|------|-------------|
| Target group | Student, Technical Staff and Academic Staff |
| Topics covered | Number of attacks faced, type of attack, loss due to attack, awareness about the security policy and reporting mechanism. |
| **Key findings** ||

- This part of the survey reveals critical facts that are related to the security threats experience of the respondents.
- 71 percentages of the respondents faced security attacks in a range of 1 to 3.
- Very surprisingly 13 percentages of the respondent are the victims, who faced attack more than 10 times in a year.

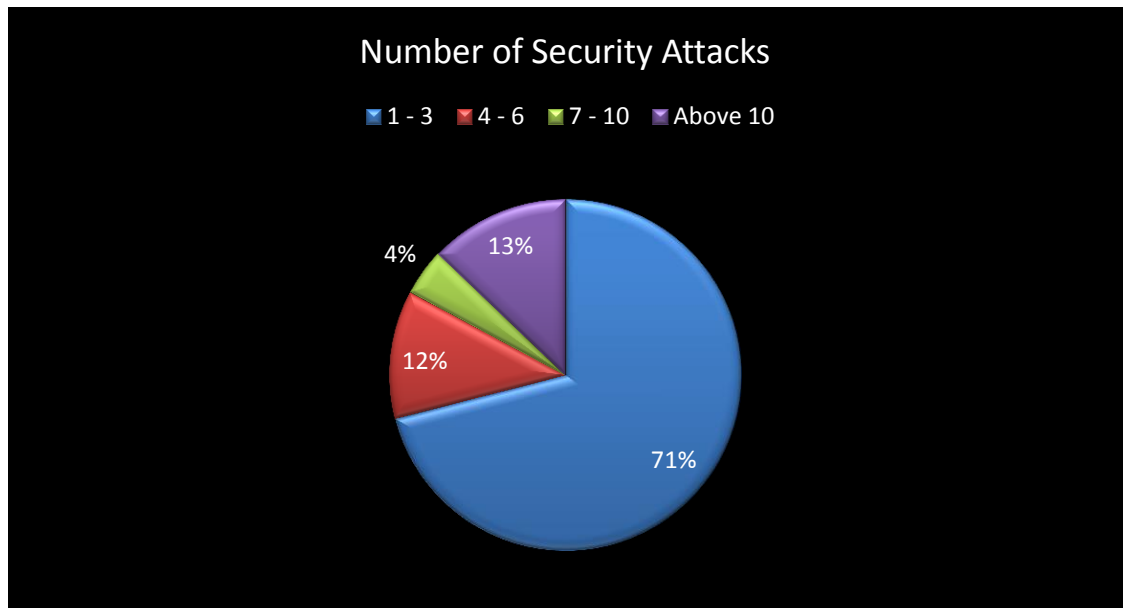

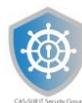

- As a result of security attacks, 36 percentages of the respondents has lost their personal data and 35 percentages of the respondent systems crashed.

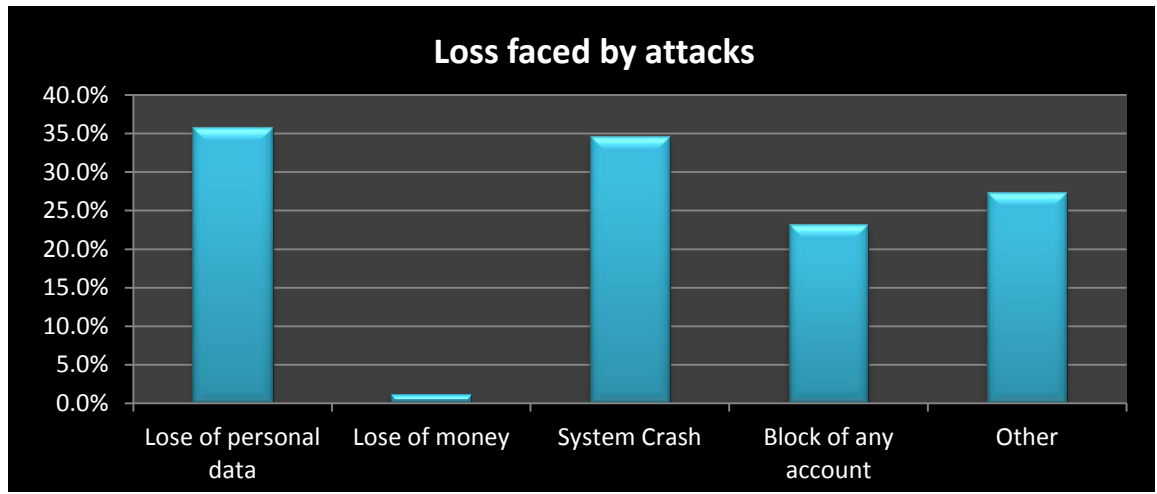

- The major types of security attack experienced by the respondents are virus, spam and Adware.

- The survey reveals a bitter truth that 39 percentages of the respondents are not aware of the security policy of their institutions and 17 percentages reveals they do not have any security policy.

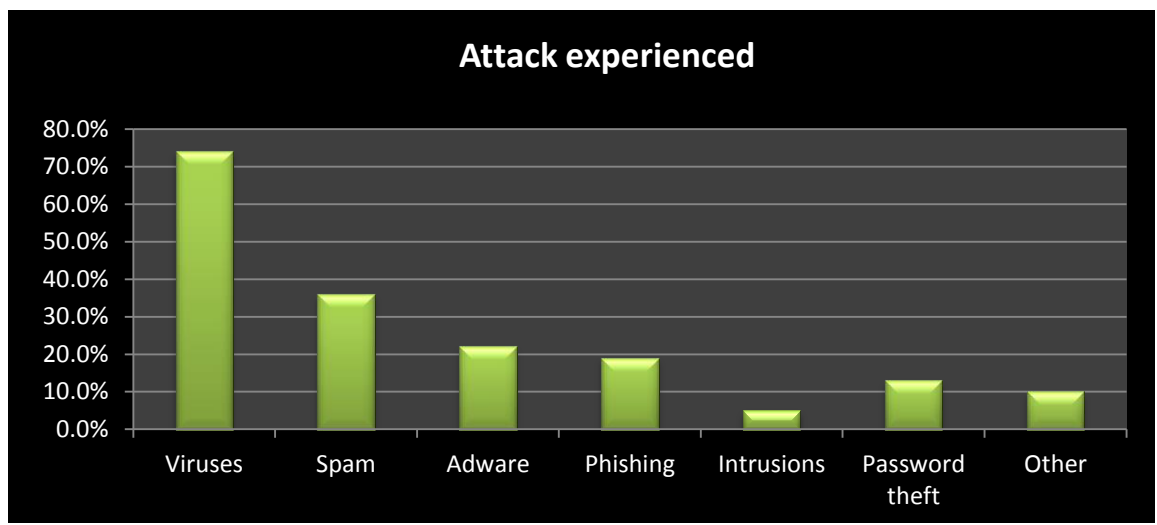

- 38 percentages of the respondent are not aware of the reporting mechanisms and 25 percentages of institutions do not have any reporting mechanism.

- Most of the respondents are not interested in reporting the security incidents.

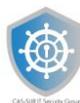

**2.4 Password awareness**

| Item | Description |
|---|---|
| Target group | Student, Technical Staff and Academic Staff |
| Topics covered | Password choosing, constructing, number of characters preferred, way of managing from disclosure and frequency of changing. |
| **Key findings** ||

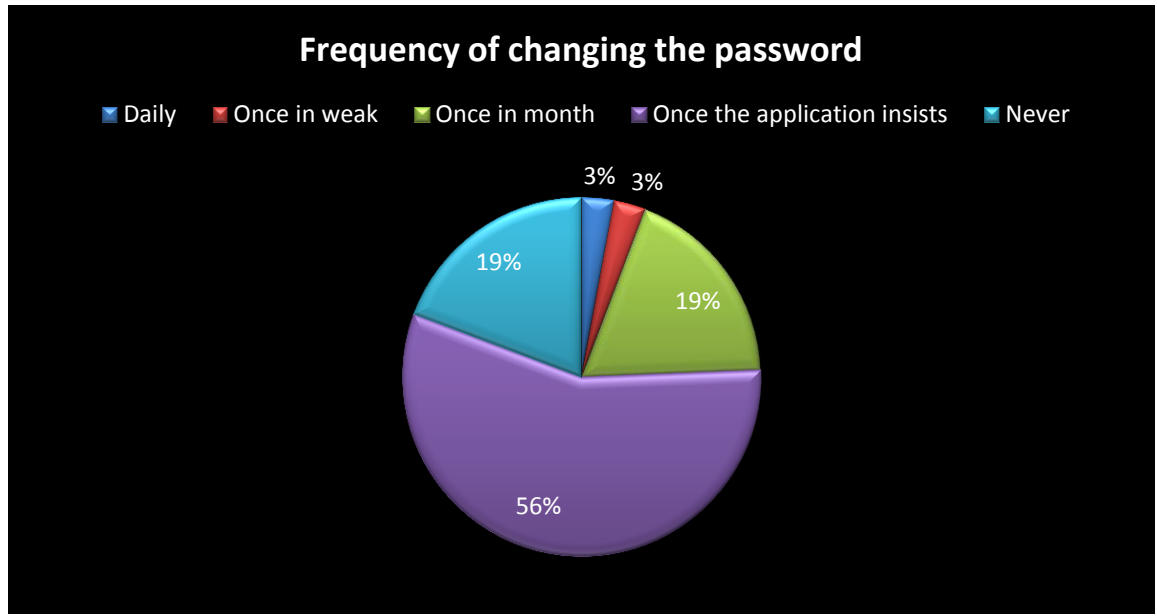

- The result shows that a wider group of respondents is having a better awareness on password management.

- Around 50 percentages of the respondents are using unique passwords for different applications, while 17 percentages of the respondents use the same password for all the applications.  Fifty percent of the respondents construct their passwords using the combination of number, alphabets and symbols. Fifty nine percentages of the respondents prefer a password with the length of 6 to 9 characters. The majority of the respondents (84 %) manage their password by memorizing.

- 78 percentages of the respondents are not willing to share their passwords with others while 7 percentages are willing to share.

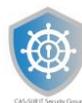

- The survey shows 19 percentages of the respondents are not at all changing their passwords and 57 percentages of the users change their password only when the application insists.

## 2.5 Content awareness

| Item | Description |
|------|-------------|
| Target group | Student, Technical Staff and Academic Staff |
| Topics covered | Response to an unknown email, opening an email attachment, email policy of organization, handling email from unknown persons, phishing emails, hoax emails and chain emails. |
| **Key findings** | |

- 32 percentages of the respondents show their interest in opening an email from the unknown source.

- 39 percentages of the respondents agreed that there is no email policy in the institution and 23 percentages revealed there are policies, but I do not know and I could not understand.

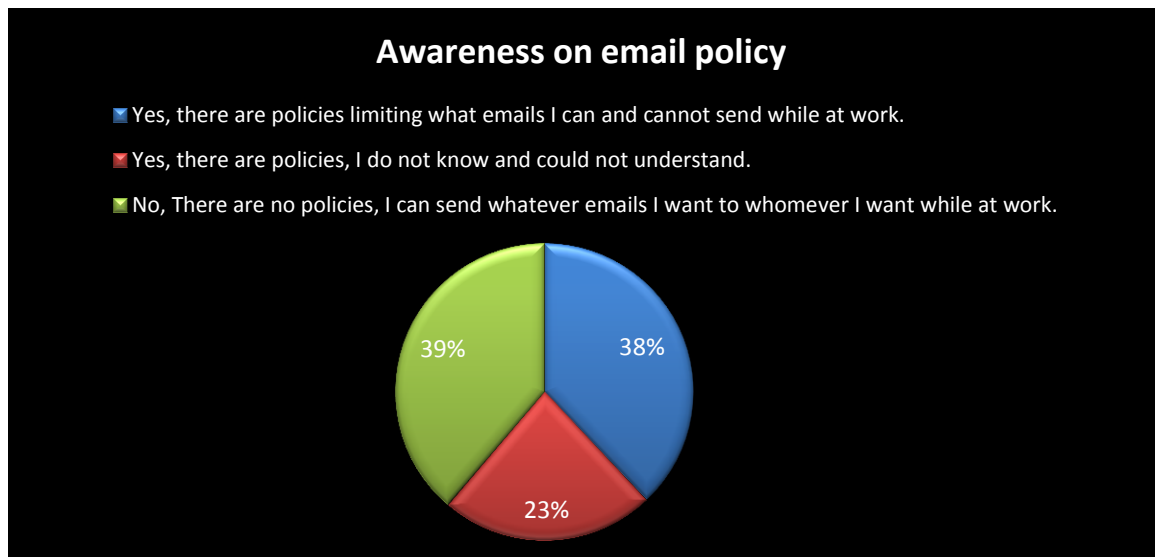

- Regarding the phishing emails, which tries to collect the personal information from the users, 84 percentages of the respondents says that they do not reveal their personal information and mere 3 percentages of the respondents are willing to provide their bank

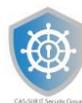

details.

- 79 percentages of the respondent are not ready to encourage hoax mails and chain mails.

## 2.6 Security practice

| Item | Description |
|------|-------------|
| Target group | Student, Technical Staff and Academic Staff |
| Topics covered | Components to develop positive security habits, confidentiality about the institution protection, security awareness program and security training programs. |
| **Key findings** | |

- The majority (64%) of the respondents thinks that the knowledge, the skills and the attitude are the necessary components to develop a positive security habits.

- Almost 40 percentages of the respondents are not confident about their institutions protection against information security risk.

- 56 percentages of the respondents exposed that their institution never conducted security awareness program.

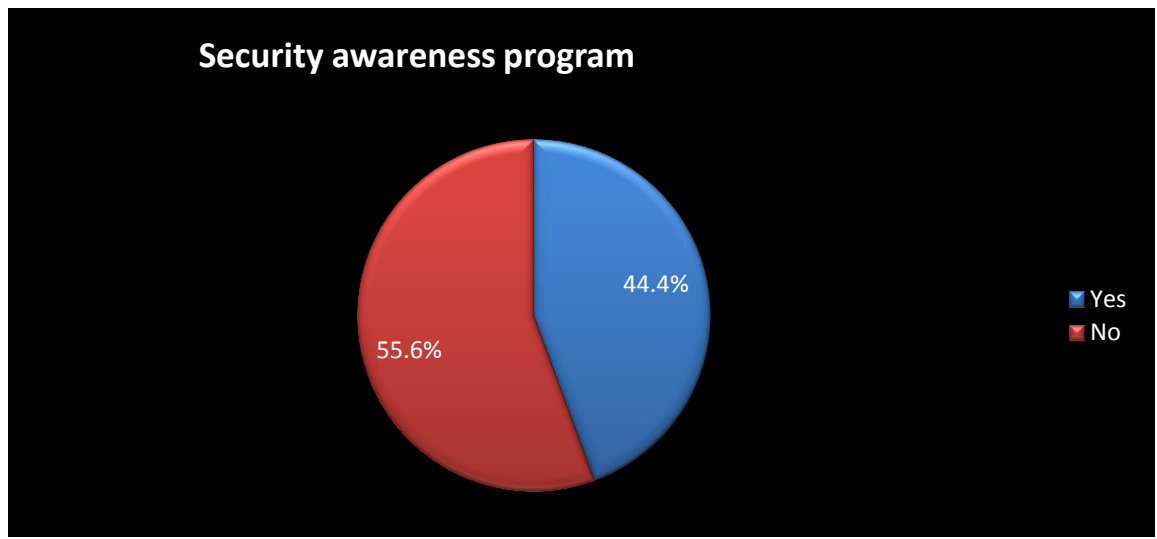

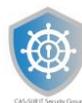

- 59 percentages of the respondents have confirmed that they have not attended training in the past 12 months.

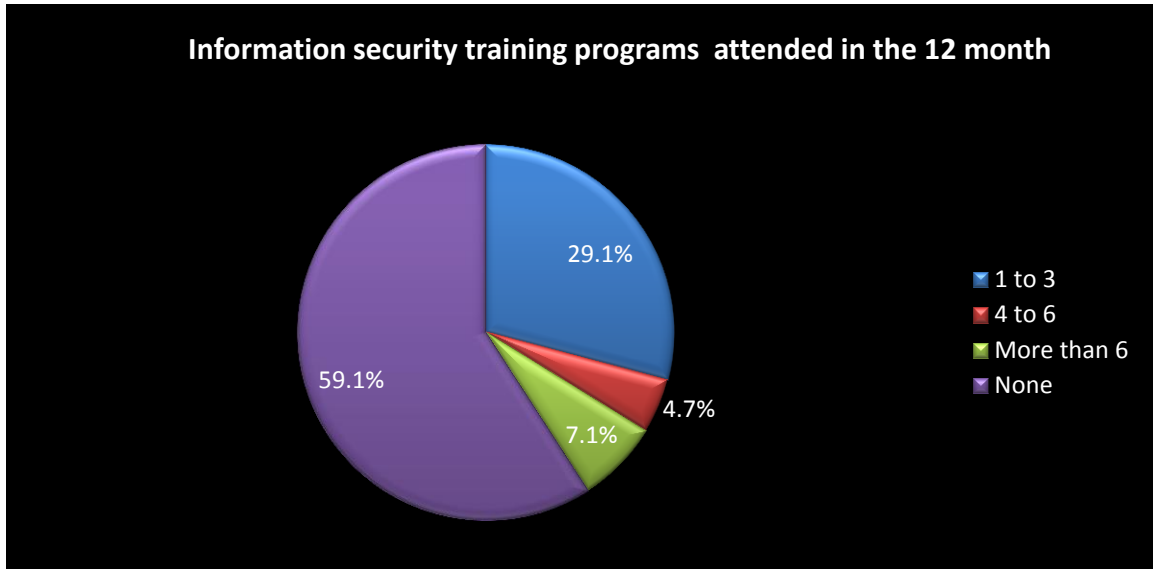

# CHAPTER 3

# TEAM WORK

Studying about information security awareness involves different stages, methods and approaches to reach the target group, collecting the survey and analyzing the response.  It becomes necessary to achieve the task with different expertise to achieve the agenda.  A team was formed to achieve the goal.

## 3.1 Team selection

The following team members with the knowledge of IT Security were identified:

| Designation | Name |
|---|---|
| Coordinator | Mr. Rajasekar Ramalingam |
| Member | Dr. Ramkumar Lakshminarayanan |
| Member | Mr. Shimaz Khan Shaik |

## 3.2 Collective skills and experience of team members

| Coordinator | Rajasekar Ramalingam |
|---|---|

*Rajasekar Ramalingam is post graduate in Computer Science and Engineering, currently working as lecture in the department of Information Technology, College of Applied Sciences, Sur, Oman.  He is having 12 years of experience in teaching.  His area of specialization is Computer Network and Security.  He has worked in various engineering colleges in India and conducted various training programs in Oman and India.  He has presented papers in national conferences, in international journals and attended many conferences and workshops.*

| Member | Ramkumar Lakshminarayanan |
|---|---|

*Ramkumar Lakshminarayanan is post graduate in Computer Science, PhD in Commerce and currently working as a Lecturer, Information Technology in College of Applied Sciences, Sur Oman. He is having 14 years of experience in teaching, consulting and software development. He has conducted training for leading corporate companies in India and abroad in the field of Database; Data ware housing, Networking, IT Security, Cloud computing, and Mobile Technology.  He has presented articles in various journals and attended conferences around the Globe.*

| Member | Shimaz Khan Shaik |
|---|---|

*Shimaz Khan Shaik is a post graduate in the Computer Applications and at present working as a Lecturer in the IT Department of Sur College of Applied Sciences, Sur, Oman. His area specialization is Networking and IT security. He is having 12 years of experience in the field of teaching, presented papers in national and international conferences.*

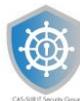

### 3.3 Team member's tasks and KPI

| Task | Assigned to | KPI |
|------|-------------|-----|
| Finding the key area to be surveyed. | Rajasekar, Ramkumar | • Setting up the Questionnaire<br>• Reaching the audience<br>• Collecting response<br>• Analyzing the response<br>• Report preparation |
| Designing the Survey questionnaire | Ramkumar, Shimaz | |
| Uploading the Survey | Shimaz, Rajasekar | |
| Reaching the target group | Rajasekar, Ramkumar | |
| Promoting the survey link | Ramkumar, Shimaz | |
| Following the responses | Shimaz, Rajasekar | |
| Survey Analysis and preparing the report | Rajasekar, Ramkumar, Shimaz | |

### 3.4 Challenges faced

Challenges faced by the team during the implementation of the survey:

- Lack of support from the target group during the initial period.
- Lack of knowledge on the local language to reach the community and others effectively.
- Difficulty in expanding the target group from IT students to Non IT students.
- Difficultly in involving the Non IT faculty members.

### 3.5 Aligning with the project strategy and objectives.

All the members were aligned to the project strategy and objectives to implement the information security awareness survey among the educational institutions in Oman.

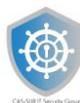

# CHAPTER 4

## FUTURE PLANS

- Look to other organization with a similar user landscape and evaluate their practice of information security awareness.
- Design a model policy and initiate the educational institutions to adopt in order to strengthen their IT infrastructure.
- Conducting future evaluation with this baseline report.
- Information security awareness is not get-it and forget-it activity, it needs periodical support and propose an awareness program specific to educational institutions in Oman.

# CHAPTER 5

## CONCLUSION

Information security awareness is an essential and a foundational element in assuring the nation's information assets are protected and always ready to meet mission objectives. The study on Information security awareness in educational institutions found several important issues that need to be addressed. Although the student, technical staff and academic staff are having the basic knowledge of security, still they are not aligned to the security practice of the institutions. There seems to be a significant lack of understanding in protecting information assets of the institutions. There needs to be a greater sense of urgency on the part of the government, other professional bodies and the educational institution to educate users about the information security needs of an institution. Implementing awareness training programs will never solve the purpose. The educational institutions must be tested and held accountable to ensure information security policies and procedures are understood and followed by their entities. This survey identified that, as an individual, the knowledge of information security awareness is considerably better but as an institution, information security awareness should be improved and there is an immediate need for security standards, policy, reporting mechanism and continuous awareness training in the educational institutions of Oman.

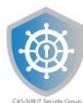

# CHAPTER 6

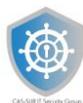